\documentclass[arxiv,usenatbib]{iupartex}  
\usepackage{newtxtext,newtxmath}
\usepackage[T1]{fontenc}
\usepackage{ae,aecompl}

\usepackage{multicol}   
\usepackage{bm}		    

\definecolor{MyDarkBlue}{rgb}{0,0.1,0.7}
\hypersetup{pdfborder={0 0 0},colorlinks,breaklinks=true,
  urlcolor={blue},citecolor={MyDarkBlue},linkcolor={magenta}}

\newcommand{\Pf}{P_{\rm f}}
\newcommand{\Pin}{P_{\rm i}}
\newcommand{\Pmin}{P_{\rm min}}
\newcommand{\Pmax}{P_{\rm max}}

\title[The Period Clustering of Magnetars and XDINS]{The Period Clustering of Magnetars and X-ray Dim Isolated Neutron Stars}

\author[Ekşi, K.Y.]{Kazım Yavuz Ekşi$^{1\cc}$\orcid{0000-0001-5999-0553}
\affsep \\
$^1$İstanbul Technical University, Faculty  of Science  and  Letters, Physics Engineering  Department, İstanbul 34469, Türkiye
}

\corres{K.~Y.~Ekşi}{eksi@itu.edu.tr}

\pubyear{2026}

\doiheader{XXXXXXX/PAR.20XX.00000}
\date{
	\pSubmit{00.00.0000} 
	\pRevReq{00.00.0000}
	\pLastRevRec{00.00.0000}
	\pAccept{00.00.0000}
	\pPubOnl{00.00.0000}
}
\volume{0}
\volnumber{1}

\begin{document}
\label{firstpage}
\pagerange{\pageref*{firstpage}--\pageref*{lastpage}}
\maketitle

\begin{abstract}
The spin periods of magnetars and X-ray dim isolated neutron stars (XDINS) cluster within a remarkably narrow range. Using the current sample of 30 magnetars with measured periods (ranging from 0.33 to 11.78~s) and 8 XDINS (ranging from 3.45 to 12.76~s), we utilize the point-likelihood technique to constrain the birth and final periods of these sources, assuming a steady-state population. Employing a general braking law characterized by a constant braking index $n$, we find that for $n > 2$ the final (cut-off) period of magnetars is constrained to $\Pf \simeq 11.8 - 12.0$~s  and XDINS to $\Pf \simeq 12.8 - 14.9$~s, at the 95 per cent confidence level, while the birth periods remains largely unconstrained for dipole spin-down ($n=3$) as in earlier work. The slight increase in the upper cutoff from $\sim$12 to $\sim$15 s over two decades of discoveries of new sources yielding a threefold increase in the known magnetar population and the extension of the minimum period to $\sim 0.33$~s strongly support a physical origin for this clustering. We discuss this result in the context of magnetic-field-decay models and fallback-disc torque-equilibrium scenarios. The combined magnetar and XDINS sample (38 sources) yields the tightest constraints on $\Pf\simeq 12.8-12.9$ s, for $n=3$, suggesting possible evolutionary connections between these populations and pointing toward a common physical mechanism that terminates the observable phase of these neutron stars at periods near 14~s.
\end{abstract}


\section{Introduction}
\label{sec:intro}

Young isolated neutron stars manifest in a remarkable diversity of observational classes, including rotationally-powered pulsars (RPPs) and several populations with unusual properties \citep{eno+19}. Among the latter are the soft gamma-ray repeaters (SGRs) that are identified through their recurrent bursts and occasional giant flares, and anomalous X-ray pulsars (AXPs), which are originally distinguished by their steady X-ray pulsations reminiscent of accreting systems, but without evidence for binary companions \citep{mer08}. The discovery of SGR-like bursts from AXPs \citep{gav+02, kas+03} and the overall similarity of their timing and spectral properties have unified these sources into a single class, the so-called \textit{magnetars} \citep{kas04,ola14}. The spin periods of magnetars are in the range of $0.33$~s to $11.79$~s and, assuming magnetic dipole spin-down, their large spin-down rates indicate to strong magnetic fields in the range of $B_{\rm d} =$ a few $10^{12}$~G to $10^{14}$~G \citep[see][for reviews]{tur+15,kas17}.

According to the magnetar model \citep{dun92, tho95, tho96}, AXPs and SGRs are neutron stars with superstrong surface magnetic fields ($B \sim 10^{14}-10^{15}$~G). The characteristic bursts result from instabilities in the crust and the magnetosphere induced by these fields. The persistent X-ray emission is powered by the decay of the magnetic fields. These superstrong magnetic fields are not necessarily in the dipole component given the existence of low-magnetic field magnetars \citep{rea+10,rea+13} with dipole fields as low as a few $10^{12}$~G as inferred from their timing properties but exhibiting the tell-tale bursts.

A parallel population of isolated neutron stars, the X-ray dim isolated neutron stars (XDINS), also known as the ``Magnificent Seven'', share several properties with magnetars \citep[see][for reviews]{hab07, tur09, kap08}. These nearby ($d \lesssim 500$~pc), thermally-emitting sources exhibit spin periods in the range $3.39-11.37$~s, magnetic fields of $B \sim 10^{13}$~G inferred from spin-down measurements, and characteristic ages of $\sim 1-4$~Myr were discovered in the ROSAT X-ray survey. Recently, a new XDINS was discovered by eROSITA with a period of 12.76 s \citep{kur+24}. Although XDINS do not show the bursting activity characteristic of magnetars, the overlap in their spin period distributions and the suggestion of evolutionary connections \citep{pop06,vig+13,alp01,ert+14} have prompted comparative studies of these populations.

A striking and long-standing puzzle is the narrow clustering of spin periods in both magnetars and XDINS relative to those of RPPs. \citet{psa02} (hereafter PM02), by employing the 10 magnetars known at that time with periods confined to the range $6-12$~s, used a point-likelihood technique to show that this clustering was statistically significant. For a dipole spin-down law ($n=3$; see Eqn.\eqref{eq:braking}), they found that the final (cutoff) period must lie very close to the maximum observed period of $\sim 12$~s, while the birth period was largely unconstrained.

In the magnetar model, an exponentially decaying magnetic field can naturally produce period clustering, as the spin-down timescale becomes comparable to the field decay timescale at periods of several seconds \citep{col+00}. In this scenario, the long-term period evolution is dictated by field decay rather than dipole radiation. \citet{pon+13} demonstrated that a highly resistive layer within the neutron star crust, located at densities where the nuclear pasta phases are expected, can limit spin periods to $\lesssim 12$~s by dissipating the magnetic field on timescales that become comparable to the characteristic age at these periods.

In the fallback disc model \citep{mic81}, isolated neutron stars are assumed to be surrounded by residual fallback material from their natal supernovae. For the magnetars, the observed periods could represent a torque equilibrium state between magnetic braking and disc accretion \citep{cha+00, alp01, ert+09}. This is \textit{not} an alternative to the magnetar model but a complementary one; the fields causing the magnetar bursts could be in the higher multipoles, and the disc could be interacting with a lower magnetic dipole field \citep{ert03,eks03}.

Over the past two decades, the known magnetar population has grown substantially. The McGill Online Magnetar Catalog \citep{ola14} now lists 30 magnetars, of which approximately 24 are confirmed with measured spin periods. Notably, the discovery of young magnetars such as Swift~J1818.0$-$1607 \citep{esp+20} with $P = 1.36$~s, and rotationally-powered objects that show magnetar bursts (PSR J1846$-$0258 with $P = 0.33$~s; PSR J1119$-$6127 with  $P = 0.41$~s) has extended the period distribution to shorter values, while the maximum period has remained at $\sim 12$~s.

In this paper, we revisit the statistical analysis of PM02 using the current magnetar \textit{and} XDINS samples. Our goals are to: (i) determine whether the period clustering remains statistically significant with the updated data and the XDINS sample; (ii) quantify the constraints on birth and final periods as functions of the braking index; and (iii) discuss the implications of these results for the physical mechanisms governing magnetar evolution.

The structure of this paper is as follows. In Section~\ref{sec:data}, we describe the current observational data for magnetars and XDINS. Section~\ref{sec:methods} presents the statistical methodology, closely following PM02. Our results are given in Section~\ref{sec:results}, and we discuss their physical implications in Section~\ref{sec:discussion}. We summarize our conclusions in Section~\ref{sec:conclusions}.


\section{Observational Data}
\label{sec:data}

\subsection{Magnetar Sample}

We compiled spin periods for magnetars from the McGill Online Magnetar Catalog\footnote{\url{http://www.physics.mcgill.ca/~pulsar/magnetar/main.html}} \citep{ola14} and recent literature. Our sample includes confirmed SGRs and AXPs with reliably measured periods, excluding candidate sources without confirmed pulsations.

Apart from the classical magnetars, we include two high magnetic field RPPs that showed magnetar-like bursts: PSR~J1846$-$0258 ($P = 0.33$~s) \citep{gav+08,liv+11} and PSR J1119$-$6127 ($P=0.41$~s) \citep{cam+00}. Note that these sources occupy a distinct region of the parameter space and may represent transitional objects between RPPs and the classical magnetars.

\begin{figure*}
\centering
\includegraphics[width=\linewidth]{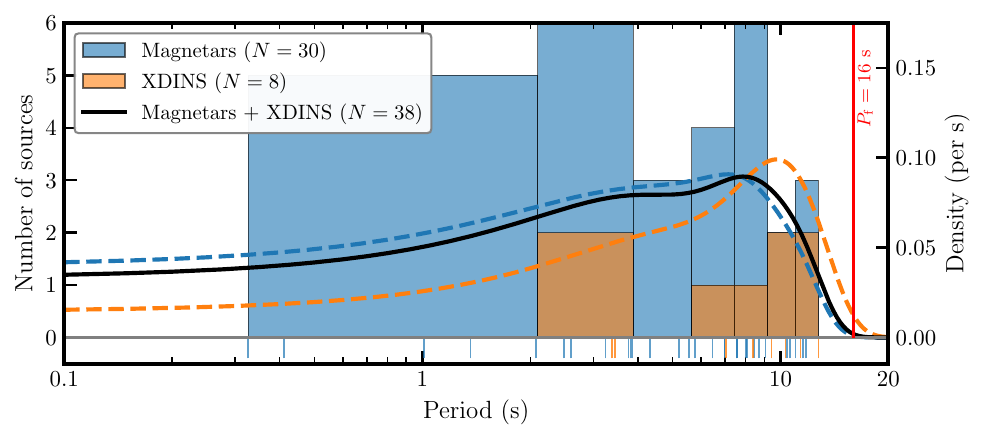}
\caption{
Spin-period distribution of the current magnetar and X-ray dim isolated neutron stars (XDINS; ``Magnificent Seven'') samples on a logarithmic axis. The blue and orange coloured histograms represent the magnetar and XDINS populations, respectively. The bin edges are obtained by Sturges' rule. Smooth kernel density estimates (KDEs) are overlaid for each population and for the total sample (38 sources; solid black curve). Short colored tick marks (``rugs'') indicate individual period measurements, with colors matching their respective populations. 
The current magnetar sample extends to shorter periods ($P \simeq 0.33$~s) than that of PM02 while the maximum period remains at $\sim 12$~s as they have found.}
\label{fig:histogram}
\end{figure*}

Our final magnetar sample comprises 30 sources with 1E~1841$-$045 having the longest period (11.78~s) \citep{dib14}. \autoref{tab:data} lists the individual sources and their periods together with the original references. For comparison, PM02 analysis used 10 magnetars with periods in the range $5.16-11.77$~s. Note also that the period of SGR 1627$-$41 was taken as $P = 6.4$~s by PM02 relying on the candidate periodicity reported by \citet{woo+99}. The updated period of this source in the McGill Magnetar Catalogue is  $P = 2.59$~s, citing \citet{esp+09}.

\begin{table}
\centering
\caption{Magnetar and XDINS sample used in this analysis. Periods are from the McGill Online Magnetar Catalog and recent literature.
[1] \citet{liv+11}; [2] \citet{cam+00}; [3] \citet{mak+16};
[4] \citet{kar+20}; [5] \citet{dib+12}; [6] \citet{kar+12}; [7] \citet{esp+09}; [8] \citet{isr+16}; [9] \citet{kas+14}; 
[10] \citet{sat+10}; [11] \citet{eno+21}; [12] \citet{lev+10}; [13] \citet{mer+06}; [14] \citet{cam+07}; [15] \citet{cam+14}; 
[16] \citet{dib+09}; [17] \citet{tor+98}; [18] \citet{dib14}; 
[19] \citet{woo+07}; [20] \citet{esp+11}; [21] \citet{mcg+05}; [22] \citet{tie+09}; 
[23] \citet{sch+14}; [24] \citet{rea+13}; [25] \citet{cot+21}; [26] \citet{an+13}; [27] \citet{rea+14}; 
[28] \citet{pir+14}; [29] \citet{hab+04}; 
[30] \citet{tie07}; [31] \citet{hab+97}; [32] \citet{zan+05}; [33] \citet{hab04}; [34] \citet{hab02}; [35] \citet{kur+24}
}
\label{tab:data}
\begin{tabular}{lcc}
\hline
\textbf{Magnetars} &  &  \\
\hline  
Source & Period (s) & Reference \\
\hline
PSR J1846$-$0258  &  0.33  & [1]  \\  
PSR J1119$-$6127  &  0.41  &  [2]   \\ 
1E 161348$-$5055  &  1.01  &  [3]   \\ 
Swift J1818.0$-$1607 & 1.36 & [4] \\  
1E 1547.0$-$5408 & 2.07 & [5] \\  
Swift J1834.9$-$0846 & 2.48 & [6] \\  
SGR 1627$-$41 & 2.59 & [7] \\   
SGR 1935$+$2154 & 3.24 & [8] \\   
SGR J1745$-$2900 & 3.76 & [9] \\  
CXOU J171405.7$-$381031 & 3.83 & [10] \\ 
Swift J1555.2$-$5402  &  3.86   &  [11]  \\  
PSR J1622$-$4950  & 4.33 & [12]  \\ 
SGR 1900$+$14 & 5.20 & [13] \\ 
XTE J1810$-$197 & 5.54 & [14] \\ 
SGR 0501$+$4516 & 5.76 & [15] \\ 
1E 1048.1$-$5937 & 6.45 & [16] \\ 
AX J1845$-$0258 & 6.97 & [17] \\ 
1E 2259$+$586 & 6.98 & [18] \\ 
SGR 1806$-$20 & 7.55 & [19] \\ 
SGR 1833$-$0832 & 7.57 & [20] \\ 
CXOU J010043.1$-$721134 & 8.02 & [21] \\ 
SGR 0526$-$66 & 8.05 & [22] \\  
SGR J1822.3$-$1606 & 8.44 & [23] \\  
4U 0142$+$61 & 8.69 & [24] \\ 
SGR 0418$+$5729 & 9.08 & [25] \\ 
SGR J1830$-$0645 & 10.42 & [26]  \\  
CXOU J164710.2$-$455216 & 10.61 & [27] \\ 
1RXS J170849.0$-$400910 & 11.01 & [18] \\ 
3XMM J185246.6$+$003317 & 11.56 & [27] \\ 
1E 1841$-$045 & 11.79 & [18] \\ 
\hline
\textbf{XDINS} & & \\
\hline  
Source & Period (s) & Reference \\ 
\hline 
RX J1605.3$+$3249 & 3.39 & [28] \\ 
RX J0420.0$-$5022 & 3.45 & [29] \\  
RX J1856.5$-$3754 & 7.06 & [30] \\  
RX J0720.4$-$3125 & 8.39 & [31] \\  
RX J2143.0$+$0654 & 9.44 & [32] \\  
RX J1308.6$+$2127 & 10.31 & [33] \\  
RX J0806.4$-$4123 & 11.37 & [34] \\  
eRASSU J131716.9$-$402647 & 12.76 & [35]  \\ 
\hline 
\end{tabular}
\end{table}

Since we include sources on the basis of exhibiting magnetar bursts, we included 1E 161348$-$5055 at the center of 2000-year-old SNR RCW~103, which does show the tell-tale bursts of magnetars \citep{rea+16, dai+16}.
The object shows modulations with a period of $6.67$~h \citep{del+06}, which is much longer than the spin period of any magnetar in the above list.  The observed modulations likely do not represent the object's spin period but rather its precession period \citep{hey02}. 
Indeed, some magnetars exhibit X-ray modulations often attributed to precession owing to their proloidal deformation due to extreme toroidal magnetic fields \citep{mak+14,mak+16,mak+21a,mak+21b,mak+24a}.
Hence, the observed variability at the $6.67$~h could be akin to the X-ray modulations observed from other magnetars, as suggested by \citet{mak+24b}.
Indeed, after the submission of the first version of this paper, evidence for a spin period of $P=1.01~{\rm s}$ was discovered by \citet{mak+26}, which is in the range of the periods of other classical magnetars in our list and hence justifies the inclusion of the source in our sample.

We exclude the ultra-long-period sources such as GLEAM-X~J162759.5$-$523504.3 \citep{hur+22} with $P \sim 18$ minutes and GPM J1839$-$10 \citep{hur+23} with $P \sim 22$ minutes. These may represent a distinct class of neutron stars or be white dwarfs \citep[see][for a review]{rea+26}.
These exotic sources lie well outside the period range of ``classical'' magnetars and XDINS, and their relationship to the populations studied here remains unclear, yet it is also possible that they could be final observable stages of magnetars with fallback discs \citep{ron+22,gen+22}.

Progenitors of neutron stars, high-mass stars, are predominantly found in binaries. The abundance of isolated neutron stars is then understood as the result of the disruption of the binary system during the supernova explosion that produced the neutron star. Only about 10 per cent of the systems remain bound and form high-mass X-ray binaries (HMXBs). Even if a magnetar forms in a binary, its magnetic field could decay rapidly over $10^4$~yrs, which is much shorter than the typical ages of HMXBs. Thus, magnetars in binary systems should be rare \citep{king19}. Yet, it could still be possible to obtain mature magnetars assuming neutron star crusts with a fortuitous combination of extremely low impurity parameters, Hall attractors, and rapid cooling \citep{igo18}.  In fact, there had been claims for the existence of accreting magnetars in binary systems \citep[e.g.][]{rei+12,fu12,eks+15} based on strong torques required to lead to the observed spin periods. Subsonic wind accretion model of \citet{sha+12} obviates the strong magnetic fields inferred for some of these systems. Magnetar bursts had been detected from the $P=0.27$~s neutron star in LS~I~$+$61$^\circ$303, a binary system \citep{torres+12} which also showed radio pulsations \citep{wen+22}.  In the following, we exclude magnetars in binary systems from our analysis, since there is only one confirmed case (LS~I~$+$61$^\circ$303) and their evolutionary paths diverge from those of isolated magnetars right at the beginning of their lives, as they remain bound to their companion \citep{suv22}.

\subsection{XDINS Sample}

The XDINS (often called the ``Magnificent Seven'') are a group of nearby, thermally emitting isolated neutron stars discovered in ROSAT data \citep{hab07, tur09}. All sources have confirmed spin period measurements from X-ray timing observations \citep{kap08, vanker08, pir+25}
ranging from 3.45~s (RX~J0420.0$-$5022) to 11.37~s (RX~J0806.4$-$4123), overlapping substantially with the magnetar period distribution. We have added to the list, the recently discovered XDINS, eRASSU J131716.9-402647, with a period of 12.76 s \citep{kur+24}. \autoref{tab:data} lists the XDINS sample with original references.

For RX J0720.4-3125, \citet{ham+17} proposed that the true spin period may actually be, $16.78$~s i.e.\ twice that of the commonly reported value \citep{hab+97}. NICER observations by \citet{bog24}, however, 
found no significant contribution to the signal power from the pulsar at
this period. We thus proceeded with the period value reported by \citet{hab+97}, but note that including this source with $P=16.78$~s would alter the maximum observed period.

\subsection{Combined Sample}

For part of our analysis, we consider a combined sample of magnetars and XDINS, comprising 38 sources with periods ranging from 0.33~s to 12.76~s. This combined analysis is motivated by the suggestion that magnetars and XDINS may share evolutionary connections \citep{pop06}, possibly representing different stages of magneto-thermal evolution for neutron stars with initially strong magnetic fields \citep{vig+13} and different modes of fallback disc-neutron star magnetosphere interaction \citep{alp01,gen24}.

\autoref{fig:histogram} shows the period distributions for the current magnetar sample (30 sources) and the XDINS sample (8 sources). The histogram, for a modest number of data points as in here, depends sensitively on arbitrary choices such as the bin width and bin boundaries. We used Sturges's rule to determine the number of bins\footnote{Sturges's rule suggests that, given $N$ samples, the number of bins must be $k = 1 + \log_2 N$, rounded to the nearest integer. We also applied Doane's rule, an improvement on Sturges' rule for skewed data, but it had no effect on the histogram. Using the Freedman-Diaconis rule, however, resulted in a flat histogram. Since we derive no conclusion from the histogram itself, we do not present each version.}.
Thus, together with the histogram, we show the kernel density estimates (KDEs) for each population. 
A KDE is a non-parametric method for estimating a continuous probability density function by replacing each data point with a smooth kernel (typically a Gaussian as in this case) and summing their contributions. A KDE produces a smooth, differentiable estimate of the underlying density that is free of binning artifacts. The figure also shows the individual period measurements at the bottom. The prominent features are as follows:

\begin{enumerate}
    \item The current magnetar sample extends to significantly shorter periods ($P_{\rm min} = 0.33$~s) compared to the original sample ($P_{\rm min} = 5.16$\,s) employed by PM02.
    \item The maximum period has changed slightly from $\Pmax \simeq 12$~s to $\Pmax \simeq 16$~s. One can notice the accumulation of objects near 13~s in the individual period measurements.
    \item The XDINS periods overlap substantially with the magnetar distribution, particularly at longer periods.
\end{enumerate}

\section{Methods}
\label{sec:methods}

We closely follow the statistical framework developed by PM02, which employs a point-likelihood technique to constrain the birth period $\Pin$ and final (cutoff) period $\Pf$ of magnetars from their observed period distribution.

\subsection{Spin evolution with constant braking index}
\label{sec:constant_n}

We assume a general braking law of the form
\begin{equation}
\dot{\Omega} = -\kappa \Omega^n,
\label{eq:braking}
\end{equation}
where $\Omega = 2\pi/P$ is the angular spin frequency, $\kappa$ is a constant proportional to the magnetic dipole moment squared (for dipole braking), and $n$ is the braking index. For magnetic dipole radiation in vacuum, $n = 3$ and $\kappa \propto B^2$, where $B$ is the surface dipole field strength.

In terms of the spin period $P$, \autoref{eq:braking} can be rewritten as
\begin{equation}
\dot{P} = \kappa' P^{2-n},
\label{eq:braking2}
\end{equation}
where $\kappa'$ is a redefined constant. The characteristic age is then $\tau_{\rm c} = P/(n-1)\dot{P}$.

We assume that all sources are born with the same initial period $\Pin$ and become undetectable when they reach a final period $\Pf$. While this is clearly a simplification --- birth periods likely have some intrinsic distribution --- it captures the essential physics and allows for tractable statistical analysis.

\begin{figure*}
\includegraphics[width=0.49\linewidth]{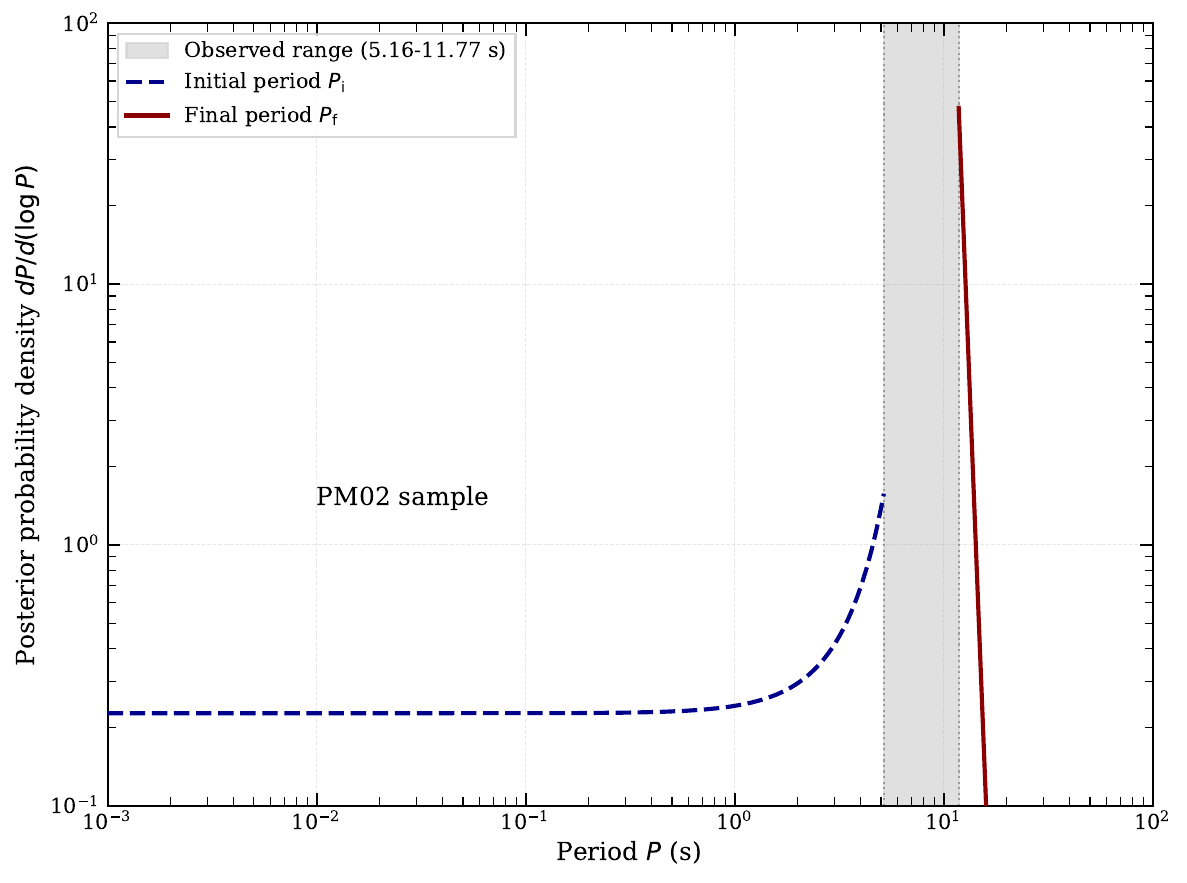}
\includegraphics[width=0.49\linewidth]{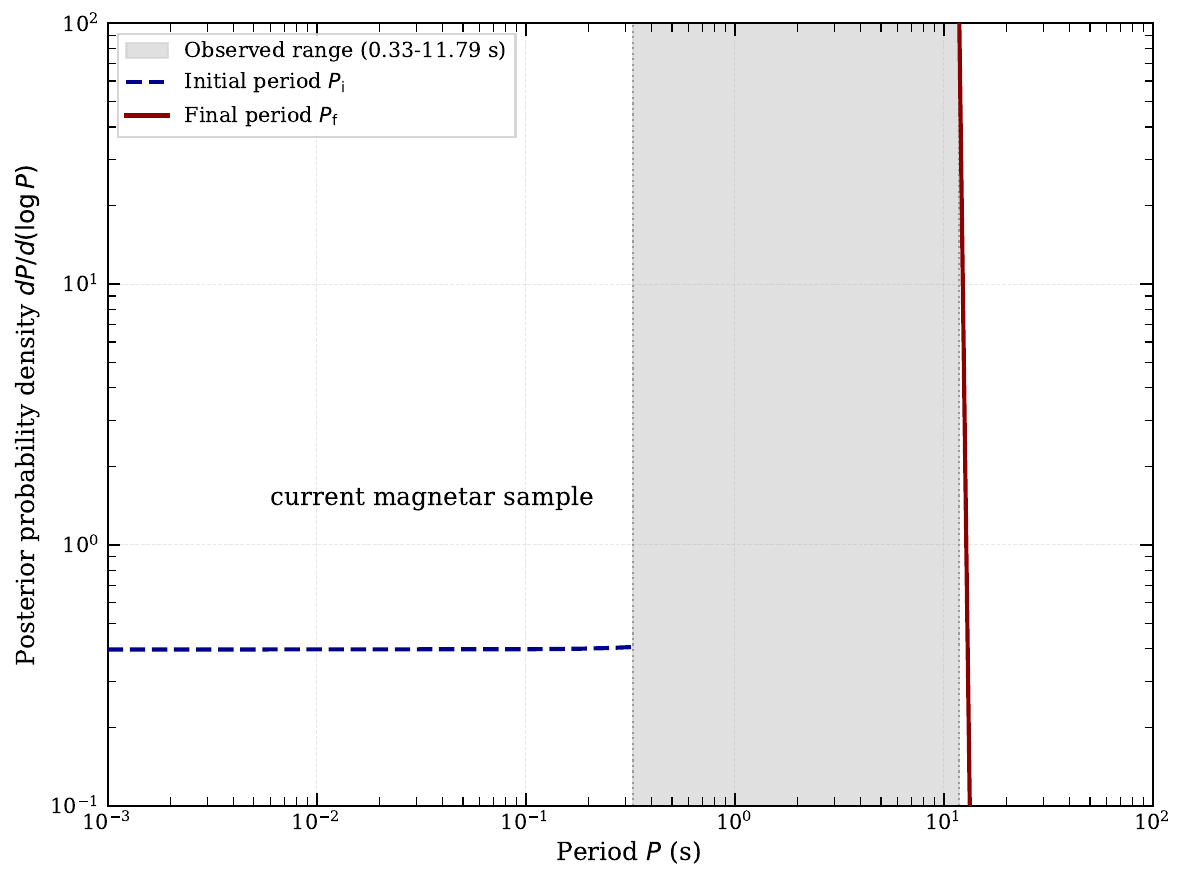}
\caption{Posterior probability distributions for the initial period $\Pin$ (dashed line) and final period $\Pf$ (solid line) for magnetars, assuming dipole spin-down ($n = 3$). The shaded region indicates the observed period range. (\textit{Left panel}) The original 10-magnetar sample to reproduce the analysis of PM02. (\textit{Right panel}) for the current sample of 30 magnetars. The broader period range ($0.33-11.78$~s) provides somewhat tighter constraints on both $\Pin$ and $\Pf$.}
\label{fig:posterior_magnetar}
\end{figure*}

\begin{figure*}
\centering
\includegraphics[width=0.49\linewidth]{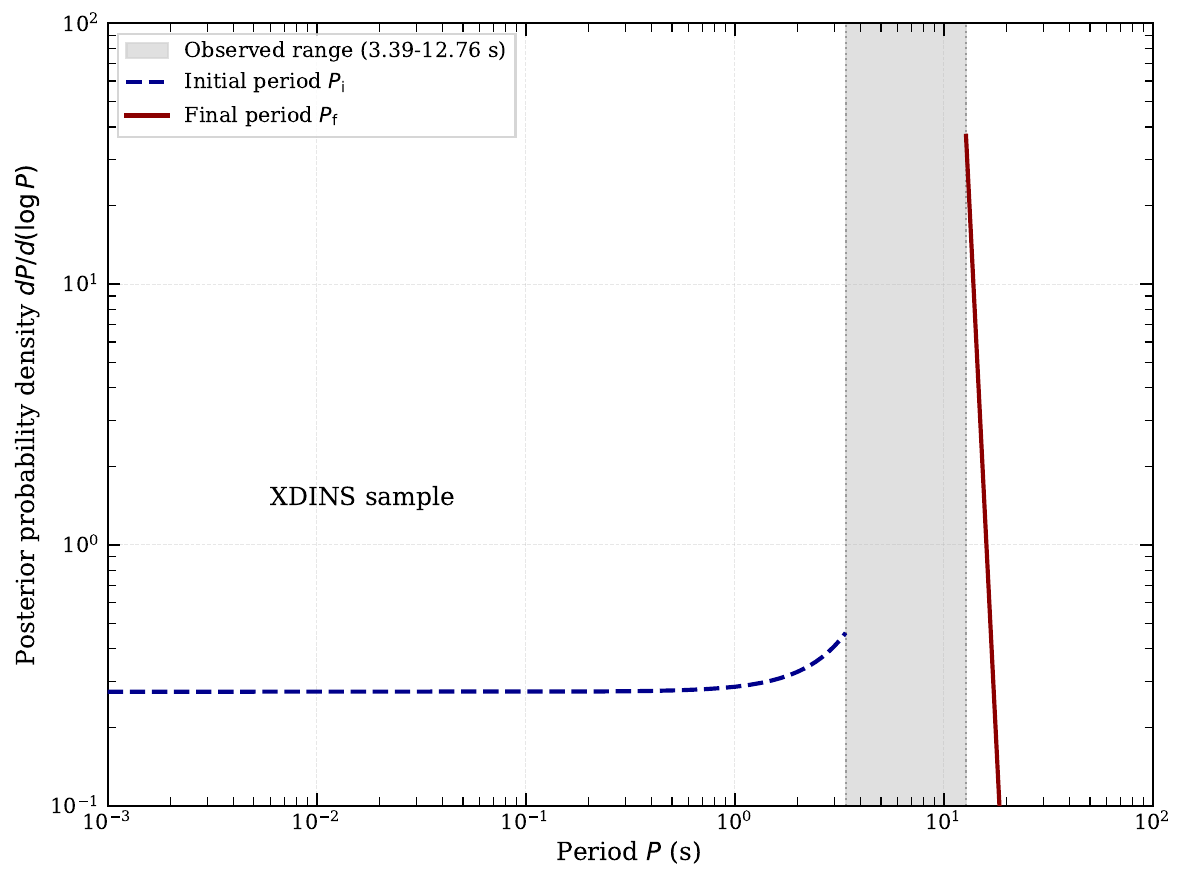}
\includegraphics[width=0.49\linewidth]{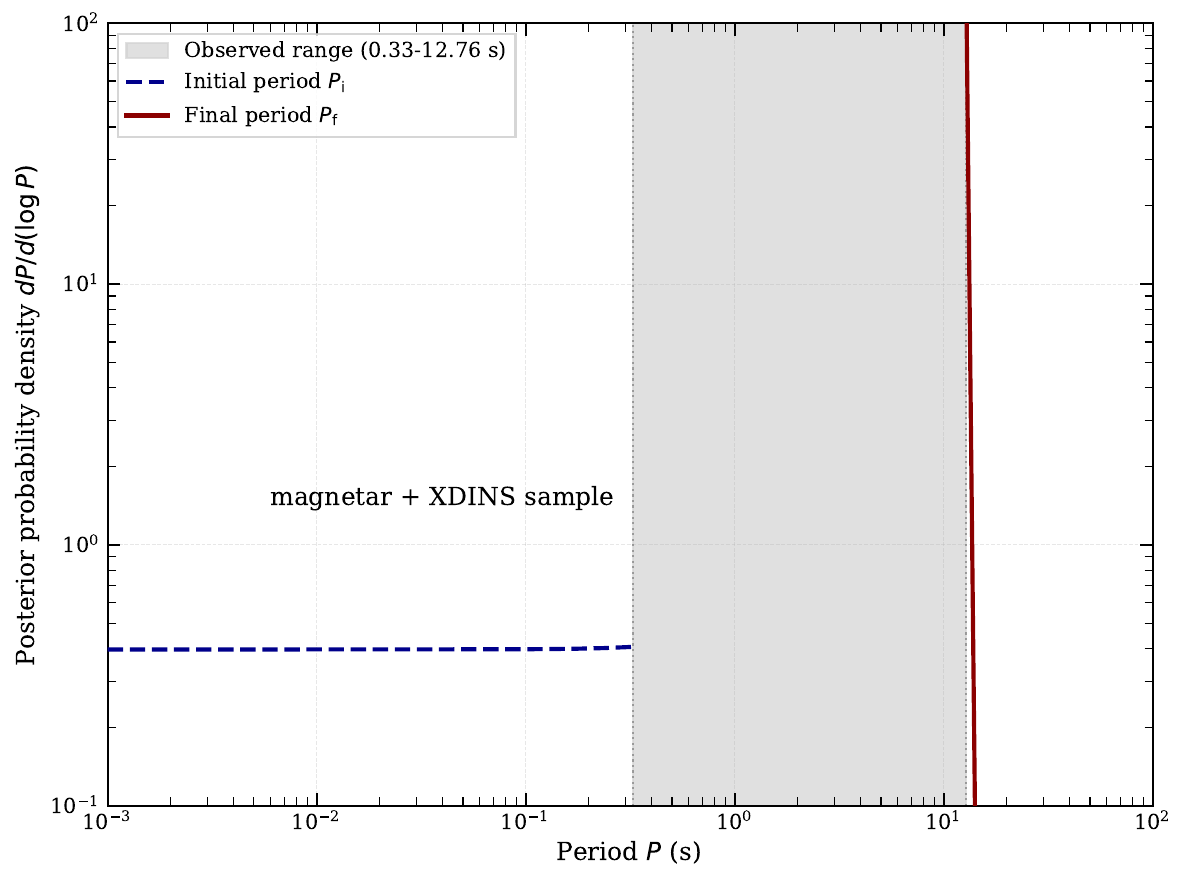}
\caption{Posterior probability distributions for the initial period $\Pin$ (dashed line) and final period $\Pf$ (solid line) for the XDINS sample (8 sources) and the combined magnetar and XDINS sample (38 sources). }
\label{fig:posterior_xdins_and_combined}
\end{figure*}

\subsection{Steady-State Period Distribution}

The evolution of the period distribution function $f(P)$ between $\Pin$ and $\Pf$ is governed by the continuity equation
\begin{equation}
\frac{\partial f(P)}{\partial t} + \frac{\partial}{\partial P}\left[\dot{P}f(P)\right] = 0,
\label{eq:continuity}
\end{equation}
where we have assumed that $\kappa$ does not evolve with time (i.e., constant magnetic field). In steady state, and using \eqref{eq:braking2} this yields
\begin{equation}
f(P) \propto \dot{P}^{-1} \propto P^{n-2}~.
\label{eq:fP}
\end{equation}
The normalized probability distribution is therefore
\begin{equation}
f(P) = C P^{n-2},
\label{eq:distribution}
\end{equation}
where the normalization constant $C$ can be determined 
by the normalization condition
\begin{equation}
\int_{\Pin}^{\Pf} f(P) \, dP = 1~.
\end{equation}
as
\begin{equation}
C = 
\begin{cases}
(n-1)/(\Pf^{n-1} - \Pin^{n-1}), & \text{ for } n \neq 1, \\
1/\ln(\Pf/\Pin) , & \text{ for } n = 1~.
\end{cases}
\label{eq:normalization}
\end{equation}
Note that the $n=1$ case of this equation differs from Eqn.(4) of PM02, correcting their typo, which we think did not affect their results.

The physical interpretation of \autoref{eq:distribution} is straightforward: for $n > 2$, sources spend more time increasingly at longer periods as they spin down, leading to $f(P)$ increasing with $P$. Conversely, for $n < 2$, sources spend more time at shorter periods.

\subsection{Likelihood analysis}

Given $m$ observed sources with measured periods $\{P_j\}$, $j = 1, \ldots, m$, the likelihood of the data given the model parameters $(\Pin, \Pf, n)$ is
\begin{equation}
\mathcal{L}(\{P_j\} | \Pin, \Pf) = \prod_{j=1}^{m} f(P_j) \Delta P = C^m (\Delta P)^m \prod_{j=1}^{m} P_j^{n-2}~.
\label{eq:likelihood}
\end{equation}
For numerical stability, we work with the log-likelihood
\begin{equation}
\ln \mathcal{L} = m \ln C + (n-2) \sum_{j=1}^{m} \ln P_j~.
\label{eq:loglikelihood}
\end{equation}
Note that for a given braking index $n$, the likelihood depends on the observed periods only through their product $\prod_j P_j$ and the range $[\Pin, \Pf]$ through the normalization $C$.

\subsection{Bayesian parameter estimation}

We adopt a Bayesian approach to estimate the posterior probability distributions for $\Pin$ and $\Pf$. Assuming prior probability distributions $\mathcal{G}(\Pin)$ and $\mathcal{G}(\Pf)$, the posterior distribution is
\begin{equation}
\mathcal{P}(\Pin, \Pf | \{P_j\}) \propto \mathcal{L}(\{P_j\} | \Pin, \Pf) \, \mathcal{G}(\Pin) \, \mathcal{G}(\Pf)~.
\label{eq:posterior}
\end{equation}
Following PM02, we adopt flat priors in $\log P$ (i.e., scale-invariant priors) for both parameters:
\begin{equation}
\mathcal{G}(\Pin) \propto \frac{1}{\Pin}, \quad \mathcal{G}(\Pf) \propto \frac{1}{\Pf}.
\end{equation}
The prior ranges are:
\begin{itemize}
    \item $\Pin$: from $10^{-3}$\,s (approximately the minimum period allowed by general relativity and the nuclear equation of state; \citealt{coo+94,don+13}) to the minimum observed period $\Pmin$.
    \item $\Pf$: from the maximum observed period $\Pmax$ to 100~s (an arbitrary but sufficiently large upper bound).
\end{itemize}
The marginalized posterior distribution for, e.g., $\Pin$ is obtained by integrating over $\Pf$
\begin{equation}
\mathcal{P}(\Pin | \{P_j\}) = \frac{\int \mathcal{L} \, \mathcal{G}(\Pin) \, \mathcal{G}(\Pf) \, d\Pf}{\int \int \mathcal{L} \, \mathcal{G}(\Pin) \, \mathcal{G}(\Pf) \, d\Pf \, d\Pin}~.
\label{eq:marginal}
\end{equation}
Using Equations~(\ref{eq:normalization}) and (\ref{eq:likelihood}), and substituting the log-flat priors, the marginalized posterior for $\Pin$ can be written as
\begin{equation}
\mathcal{P}(\Pin) \propto \frac{1}{\Pin} \int_{\Pmax}^{P_{\rm f,max}} \frac{1}{\Pf} \left(\Pf^{n-1} - \Pin^{n-1}\right)^{-m} d\Pf~,
\label{eq:marginal_Pin}
\end{equation}
for $n \neq 1$, where $P_{\rm f,max} = 100$~s is the upper bound on $\Pf$.

Similarly, the marginalized posterior for $\Pf$ is
\begin{equation}
\mathcal{P}(\Pf) \propto \frac{1}{\Pf} \int_{P_{\rm i,min}}^{\Pmin} \frac{1}{\Pin} \left(\Pf^{n-1} - \Pin^{n-1}\right)^{-m} d\Pin~,
\label{eq:marginal_Pf}
\end{equation}
where $P_{\rm i,min} = 10^{-3}$~s.

\subsection{Confidence intervals}

From the marginalized posterior distributions, we compute confidence intervals by integrating the posterior probability. For a parameter $x$ with posterior $\mathcal{P}(x)$, the $\alpha$ per cent confidence interval $[x_{\rm lower}, x_{\rm upper}]$ is defined such that
\begin{equation}
\int_{-\infty}^{x_{\rm lower}} \mathcal{P}(x) \, dx = \int_{x_{\rm upper}}^{\infty} \mathcal{P}(x) \, dx = \frac{1-\alpha/100}{2}
\end{equation}
where the notation with  $-\infty$ and $+\infty$ is standard in statistics for writing generic formulas that apply to any bounded or unbounded parameter space. In our case, since periods have physical bounds, the integrals are actually computed over $[P_{\min}, x_{\rm lower}]$ and $[x_{\rm upper}, P_{\max}]$.
We compute 68, 90, and 95 per cent confidence intervals for both $\Pin$ and $\Pf$ as functions of the braking index $n$.

\subsection{Numerical implementation}

The integrals in Equations~(\ref{eq:marginal_Pin}) and (\ref{eq:marginal_Pf}) are evaluated numerically using adaptive quadrature (the \texttt{scipy.integrate.quad} function in Python). The posterior distributions are computed on logarithmically spaced grids with $200-500$ points, and confidence intervals are obtained by evaluating the cumulative distribution function and interpolating.

For some combinations of parameters (particularly $n\simeq 1$), the integrals can become numerically challenging. We handle these cases by appropriate limiting expressions and careful error handling.

\begin{figure*}
\centering
\includegraphics[width=0.49\linewidth]{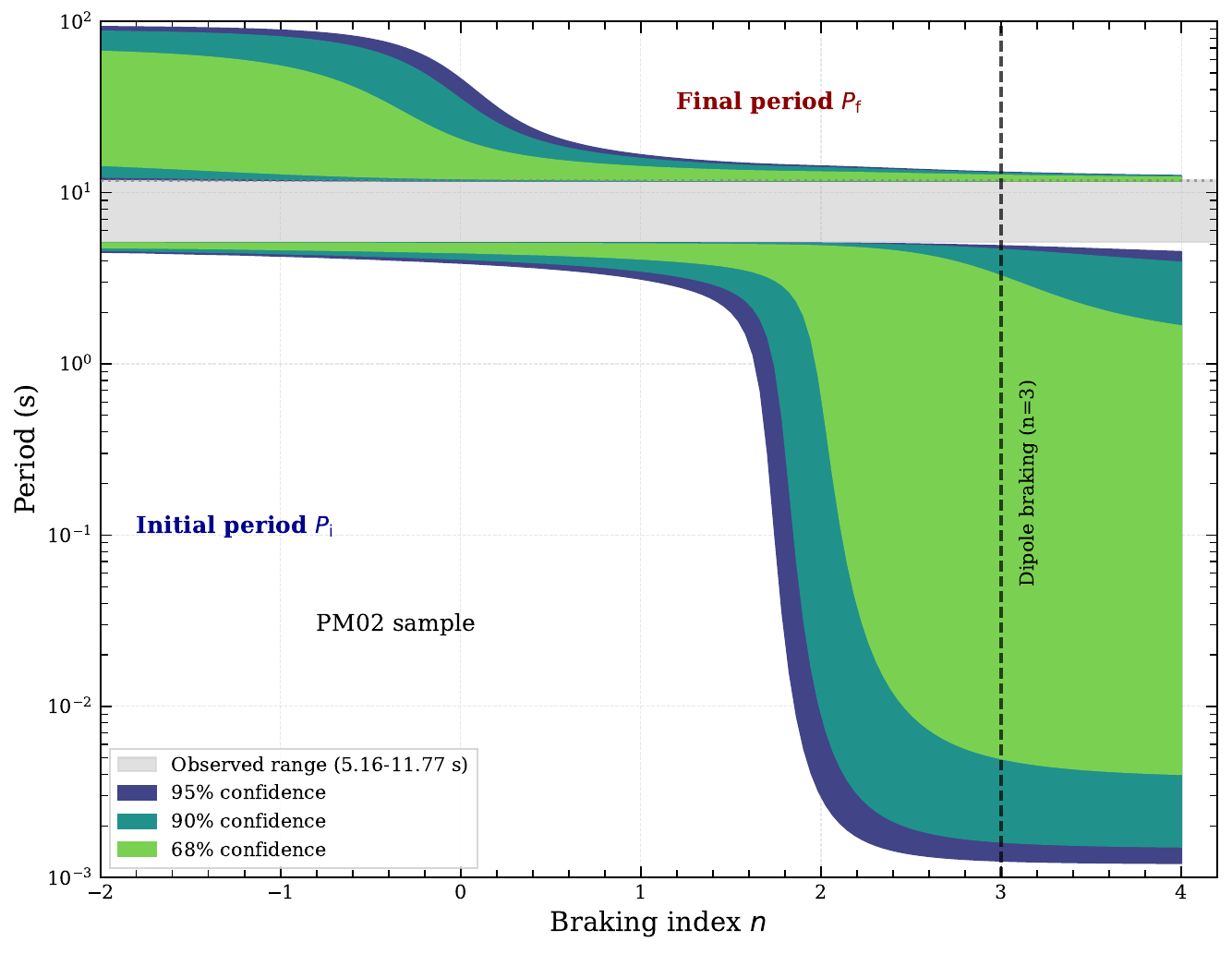}
\includegraphics[width=0.49\linewidth]{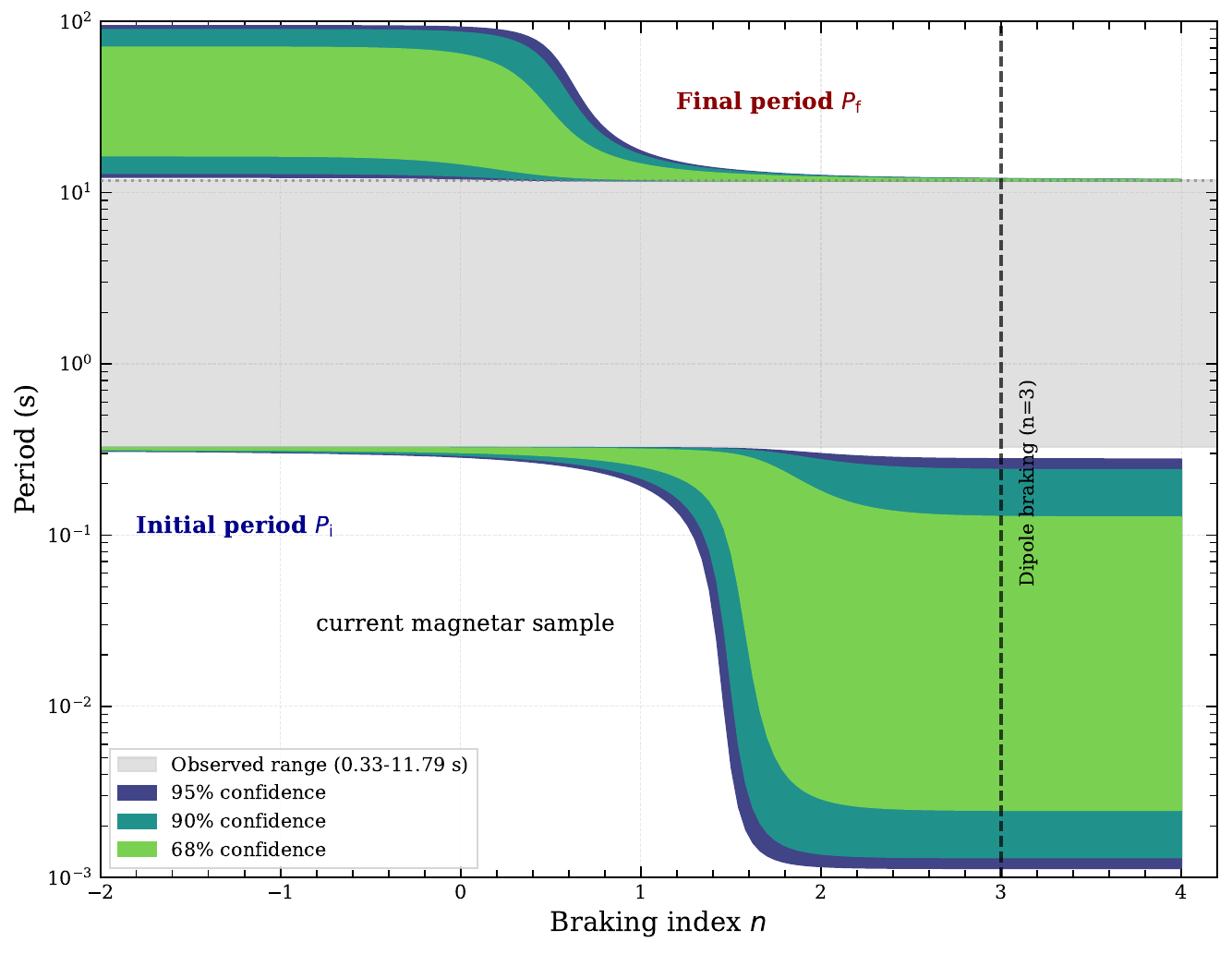}
\caption{Confidence levels (68, 90, and 95 per cent) for the initial period $\Pin$ (lower region) and final period $\Pf$ (upper region) as functions of the braking index $n$. The horizontal grey band indicates the observed period range. The vertical dashed line marks the dipole value $n = 3$. (\textit{Left panel}) is for the original 10-magnetar sample of PM02 to reproduce their Figure~2. (\textit{Right panel}) is for the current sample of 30 magnetars. The extended period range to shorter values provides tighter constraints on $\Pin$ for $n > 2$.}
\label{fig:confidence_magnetar}
\end{figure*}

\begin{figure*}
\centering
\includegraphics[width=0.49\linewidth]{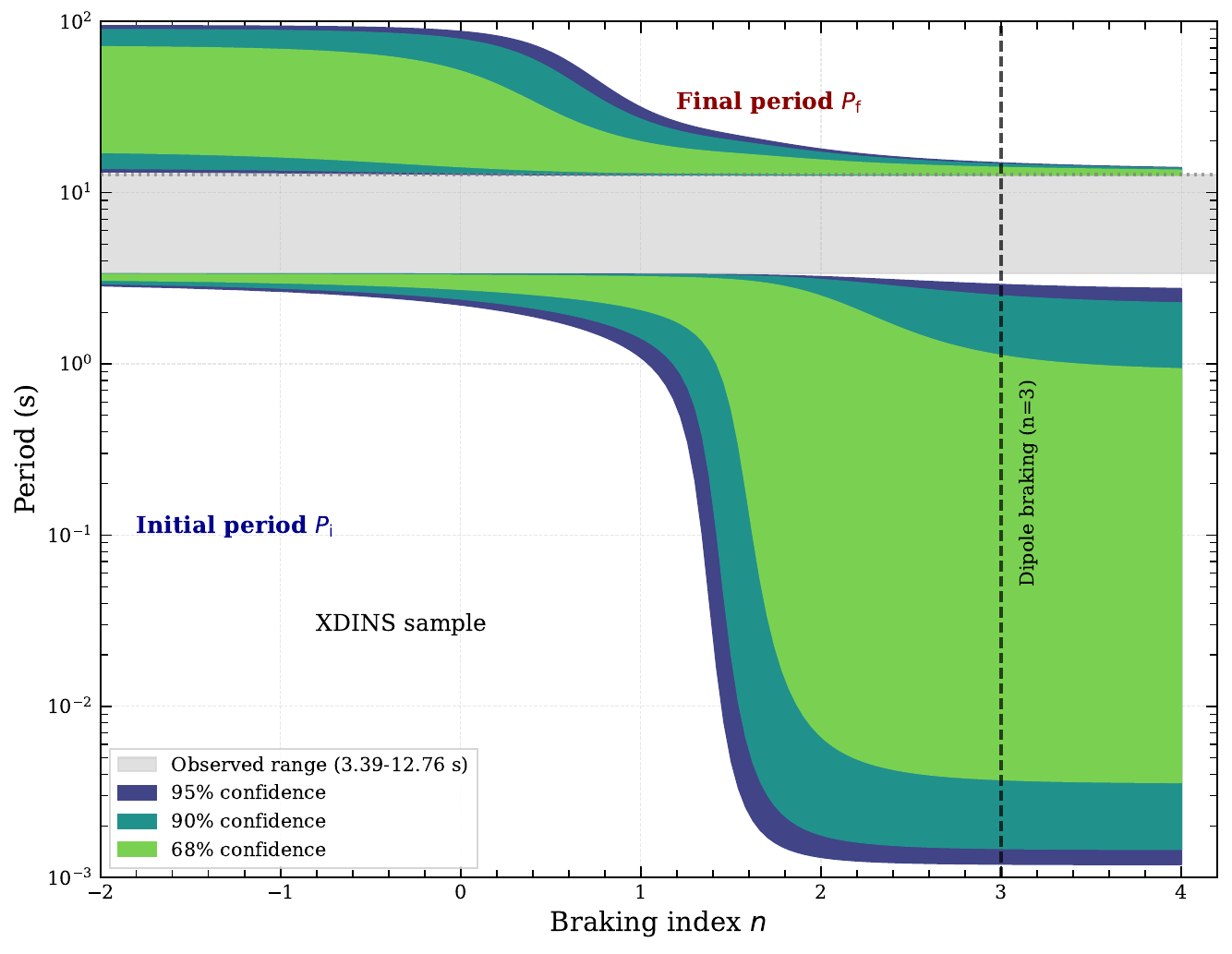}
\includegraphics[width=0.49\linewidth]{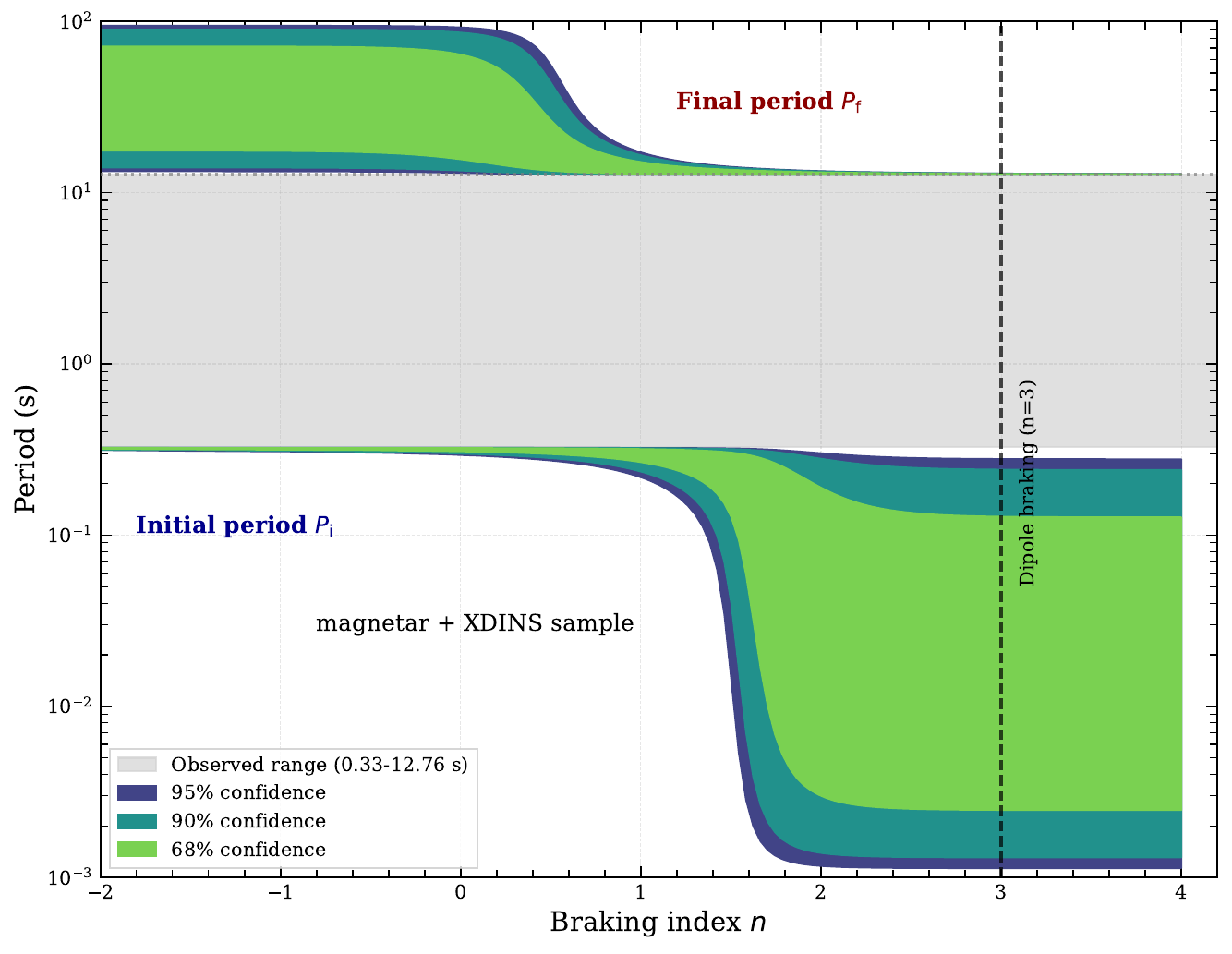}
\caption{Same as \autoref{fig:confidence_magnetar}, but for the XDINS sample (\textit{left panel}) and the combined magnetar and XDINS sample (\textit{right panel}).}
\label{fig:confidence_xdins_and_combined}
\end{figure*}


\section{Results}
\label{sec:results}

\subsection{Posterior distributions for dipole spin-down}

\autoref{fig:posterior_magnetar} shows the marginalized posterior probability distributions for $\Pin$ and $\Pf$ assuming dipole spin-down ($n = 3$), computed for the magnetars. The left panel is for the original 10-magnetar sample employed in PM02 to reproduce their Figure~1. 
The right panel is for the current sample of 30 magnetars. Note that we follow PM02 in that the plots show the distributions normalized in the $\log P$-space i.e.\  $\int {\cal P}(\log P) \, d{\log P}=1$.
The key features are:
\begin{enumerate}
    \item The posterior for $\Pf$ is sharply peaked near the maximum observed period ($\Pmax = 11.77$~s), indicating that the final period is tightly constrained.
    \item The posterior for $\Pin$ is broad and relatively flat over many decades, indicating that the birth period is poorly constrained for $n = 3$.
\end{enumerate}

\autoref{fig:posterior_xdins_and_combined} shows the corresponding posterior distributions for the XDINS sample (left panel) and the combined sample (right panel). Despite the smaller sample size, the posterior for $\Pf$ remains sharply peaked near 13~s for XDINS.
The combined sample provides the tightest constraints on the final period.

\autoref{tab:constraints} gives the 68 and 95 per cent confidence intervals for $\Pin$ and $\Pf$ for each sample, assuming $n = 3$.

\begin{table*}
\centering
\caption{Constraints on birth and final periods for dipole spin-down ($n=3$).}
\label{tab:constraints}
\begin{tabular}{lcccc}
\hline
Sample & \multicolumn{2}{c}{$\Pin$ (s)} & \multicolumn{2}{c}{$\Pf$ (s)} \\
 & 68\% CL & 95\% CL & 68\% CL & 95\% CL \\
\hline
Original & 0.005--3.25 & 0.001--4.9 & 11.8--12.6 & 11.8--13.1 \\
Current & 0.003--0.13 & 0.001--0.28 & 11.8--11.9 & 11.8--12.0 \\
XDINS & 0.004--1.1 & 0.001--2.9 & 12.8--14.0 & 12.8--14.9 \\
Combined & 0.003--0.128 & 0.001--0.278 & 12.76--12.86 & 12.76--12.92 \\
\hline
\end{tabular}
\end{table*}

\subsection{Dependence on braking index}

\autoref{fig:confidence_magnetar} shows the 68, 90, and 95 per cent confidence levels for $\Pin$ and $\Pf$ as functions of the braking index $n$. The left panel is for the original 10-magnetar sample of PM02 to reproduce their Figure~2. The key features, which were identified by PM02, are as follows:
\begin{enumerate}
    \item For $n > 2$, the final period $\Pf$ is remarkably close to the maximum observed period, while $\Pin$ is poorly constrained.
    \item For $n < 2$,  the initial period $\Pin$ is poorly constrained to relatively long values of $\sim 1-5$~s, while $\Pf$ is totally unconstrained.
    \item The transition occuring near $n = 2$ corresponds to the case where $f(P) \propto P^0 = {\rm const}$, i.e., a uniform distribution in $P$.
\end{enumerate}
\autoref{fig:confidence_xdins_and_combined} shows the corresponding plots for the XDINS sample and the combined sample.
The observed range of XDINS is much narrower than that of magnetars since the magnetar sample includes the two RPPs (PSR J1846$-$0258 and PSR J1119$-$6127), but it would be very similar if they were excluded. Again, for $n > 2$, the final period $\Pf$ is constrained to lie very close to the maximum observed period (left panel), though not as tightly as magnetars. Yet, the tightest constraint is obtained when the two samples are combined (right panel).

\subsection{Dependence on sample size}

\begin{figure*}
\centering
\includegraphics[width=0.49\linewidth]{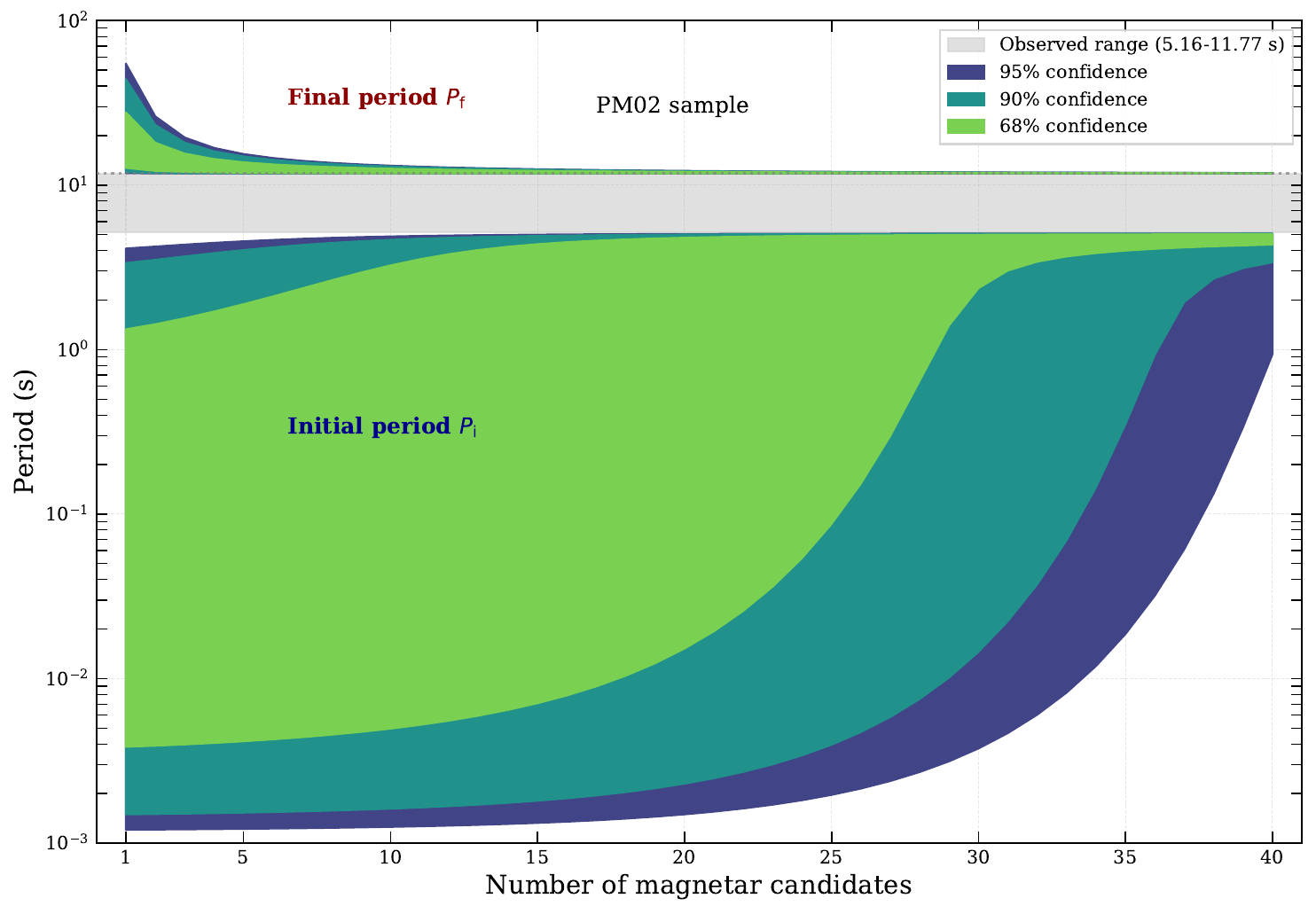}
\includegraphics[width=0.49\linewidth]{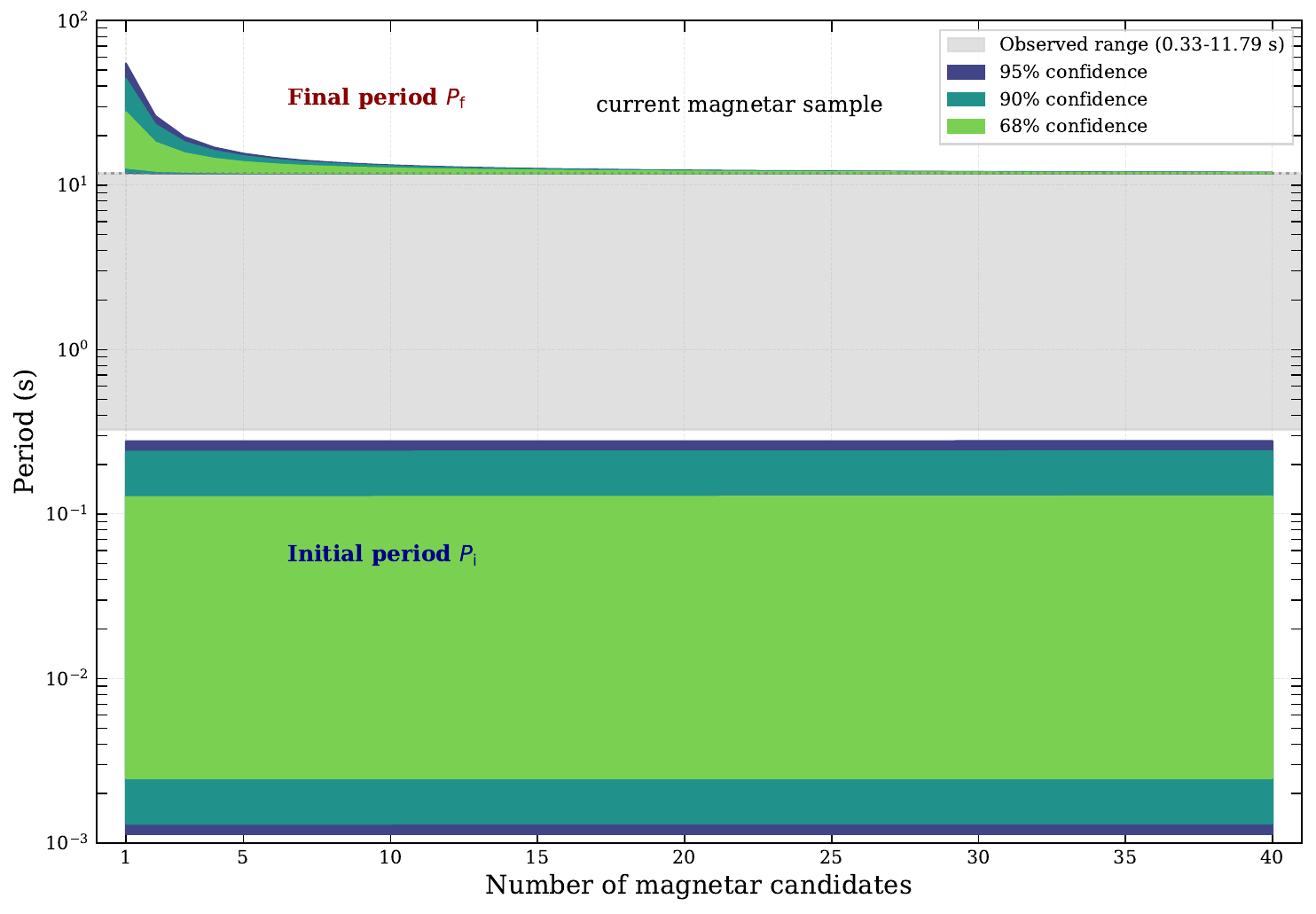}
\includegraphics[width=0.49\linewidth]{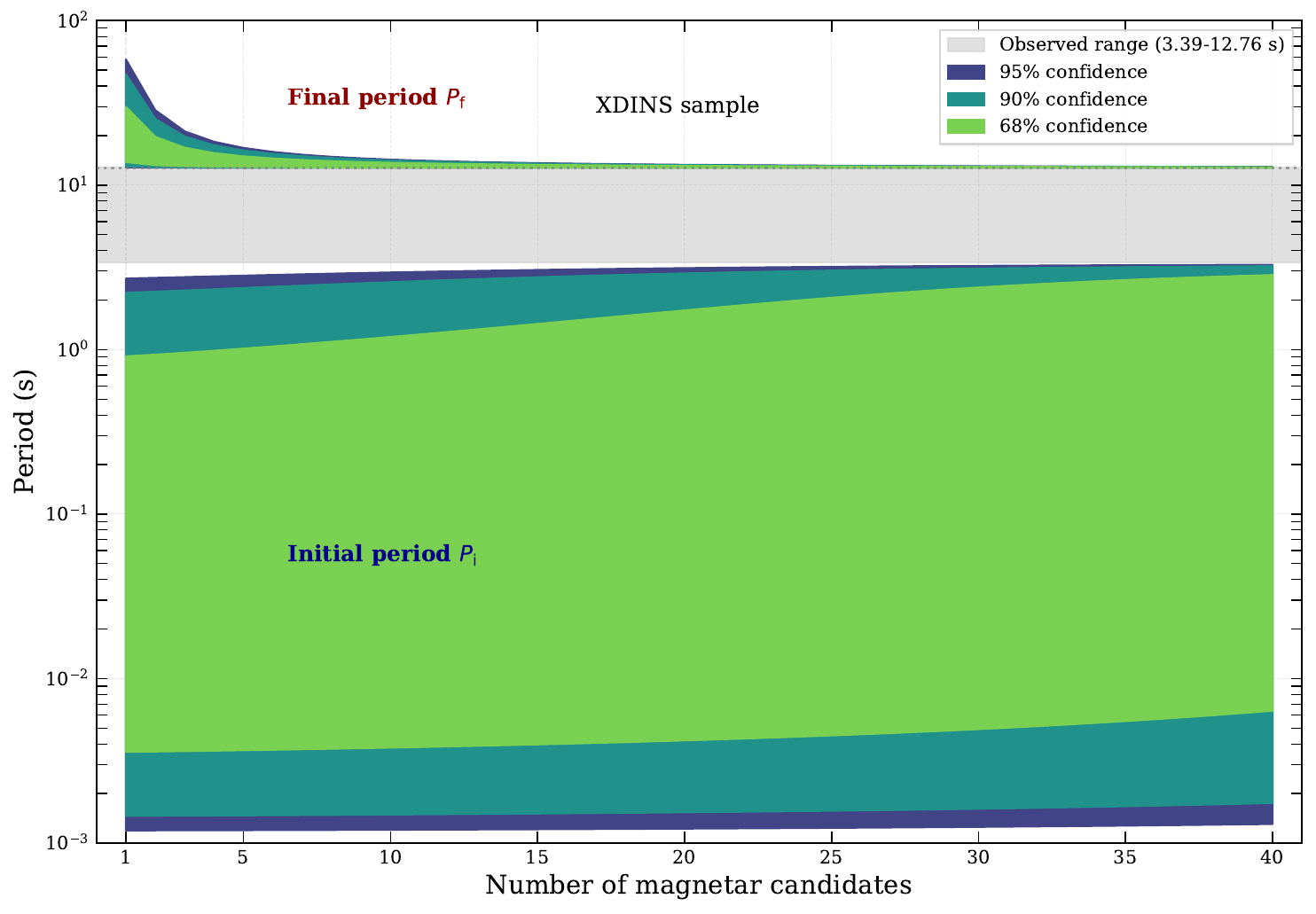}
\includegraphics[width=0.49\linewidth]{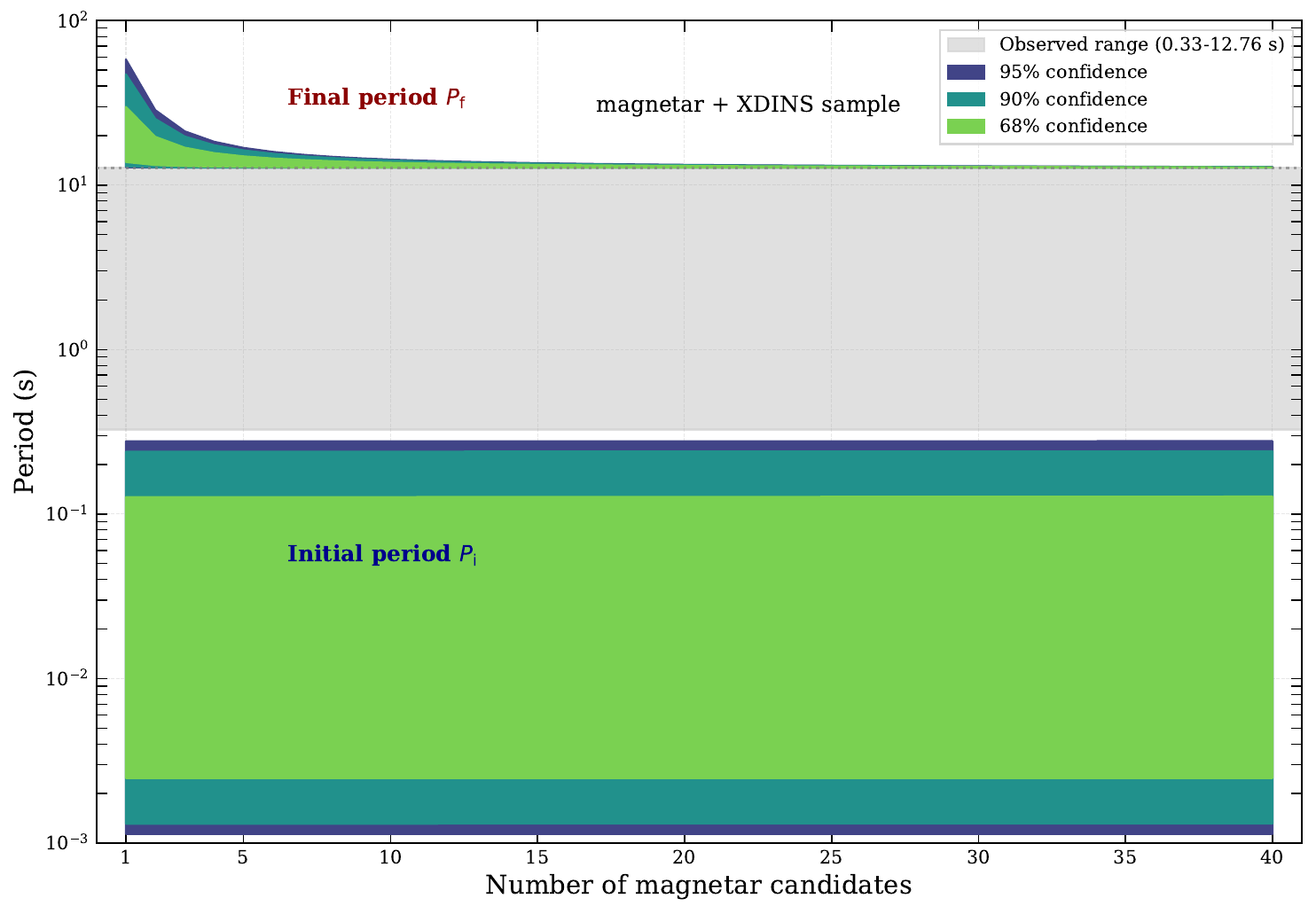}
\caption{Dependence of the period constraints on the number of sources in the sample, for dipole spin-down ($n = 3$). Markers show the 68, 90, and 95 percent confidence bounds for $\Pin$ (lower) and $\Pf$ (upper) across subsamples of different sizes. The horizontal grey band indicates the observed period range for the full sample.}
\label{fig:number_dependence}
\end{figure*}

\autoref{fig:number_dependence} examines how the constraints depend on the number of sources in each sample we considered, using the dipole braking law. Each panel shows the effect of sample size, assuming the period range remains constant.

The figure demonstrates that even with 6 sources, the constraint on $\Pf$ is quite robust, as pointed out by PM02. The constraint on $\Pin$ tightens more significantly with increasing sample size, primarily because the larger samples include sources at shorter periods.

\subsection{The evolution on the $P-\dot{P}$ derivative diagram}

\begin{figure}
\centering
\includegraphics[width=0.99\linewidth]{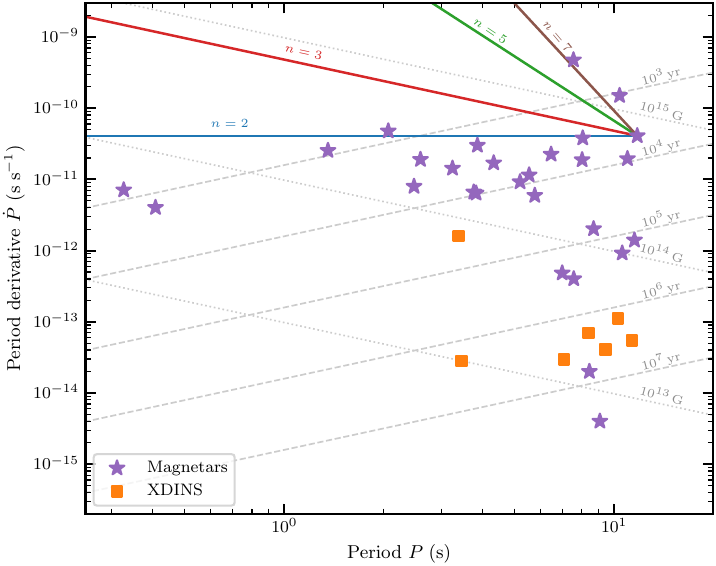}
\caption{Constant braking index tracks on the period derivative vs.\ period diagram for magnetars and XDINSs. 
Grey dotted lines are constant-$B$ loci ($10^{13}-10^{15}$~G); grey dashed lines are constant $\tau_{\rm c}$ ($10^3-10^7$ yr).}
\label{fig:PPdot}
\end{figure}

In \autoref{fig:PPdot} we show the evolutionary paths on the  $P-\dot{P}$ diagram for four braking-index tracks ($n=2,\,3,\,5,\,7$).
All tracks are anchored at 1E 1841$-$045, the longest-period magnetar in the sample, 
and drawn only for $P \leq P_0$, illustrating the evolutionary histories consistent with that endpoint. Also shown on the plot are the equi-magnetic field lines $B = 3.2\times 10^{19} \sqrt{P\dot{P}}$ and the equi-characteristic age lines $\tau_{\rm c} \equiv P/2\dot{P}$.

\section{Discussion}
\label{sec:discussion}

The most striking result of our analysis is the persistence of the $\sim 12$~s upper cutoff on magnetar and the remarkably close cutoff on XDINS periods at $\sim 16$. Despite a threefold increase in the known magnetar population over two decades, and the extension of the minimum period from $\sim 5$~s to $\sim 0.33$~s, no AXP/SGR has been discovered with a period exceeding $\sim 12$~s. Since XDINS are an order of magnitude older than AXP/SGRs, their final periods are expected to reach larger values if they are genetically connected.

For dipole spin-down ($n = 3$), we find $\Pf = 12.76 - 12.92$~s at 95 per cent confidence for the combined sample. This is remarkably close to the maximum observed period of 12.76~s, strongly suggesting that sources cease to be detectable (in X-rays) at or near this period.

This result is robust against changes in sample size (see \autoref{fig:number_dependence}) and is consistent between the magnetar and XDINS populations. 

For dipole spin-down, our analysis shows that the birth period $\Pin$ is poorly constrained: the 95 per cent confidence interval spans $10^{-3}-1$~s for the current magnetar sample. This is a direct consequence of the $f(P) \propto P$ distribution for $n = 3$: sources spend very little time at short periods, so even if they are born spinning rapidly, this leaves little imprint on the observed distribution.

However, if the braking index is $n \lesssim 2$, the constraints on $\Pin$ become much tighter. For $n = 1.5$, for example, the 95 per cent confidence upper limit on $\Pin$ is $\sim 3$~s. This is interesting because measured braking indices for young pulsars span the range $n\simeq 1.8-2.8$ \citep{liv+07, esp+17}, bracketing the critical value $n = 2$.

Magnetars themselves show highly variable and often anomalously large instantaneous braking indices ($n\sim 10^2-10^3$; \citealt{kas+00, woo+00}), likely due to glitches \citep{alp06} or fallback disc torques and timing noise. The \textit{average} braking index over the source lifetime is therefore uncertain \citep{gao+16}. If magnetic-field decay is important, the effective braking index can exceed 3.

Our analysis shows that the XDINS period distribution is statistically consistent with the magnetar distribution, supporting the hypothesis that these populations are related. The similarity of the constraints on $\Pf$ suggests that the same physical mechanism --- whether field decay, crustal resistivity, or disc equilibrium --- may be operating in both populations.

The XDINS have somewhat lower inferred magnetic fields ($B \sim 10^{13}$~G versus $B \sim 10^{14}-10^{15}$~G for magnetars), longer characteristic ages ($\tau_c \sim 1-4$~Myr versus $\tau_c \sim 1-100$~kyr), and show no bursting activity. These differences could reflect evolutionary effects: XDINS may be older magnetars whose fields have decayed to the point where bursting is no longer possible \citep{vig+13}.

Alternatively, XDINS may have been born with somewhat weaker fields than magnetars, placing them on a different evolutionary track from the outset. Population synthesis studies \citep{pop06} suggest that the birth rate of XDINS-like objects may exceed that of radio pulsars in the solar neighbourhood, implying that they represent a significant channel of neutron star formation.

The widely accepted model for the period clustering of magnetars and XDINS is that the decay of magnetic fields also sustains the persistent X-ray emission.
\citet{col+00} proposed that an exponentially decaying magnetic field can naturally explain the period clustering of magnetars. In this scenario, the spin-down rate is
\begin{equation}
\dot{P} \propto B(t)^2 P^{-1} \propto B_0^2 {\rm e}^{-2t/\tau_B} P^{-1},
\end{equation}
where $\tau_B$ is the field decay timescale. As the field decays, spin-down slows, and sources accumulate at long periods. However, when $\tau_B$ becomes comparable to the spin-down timescale $P/\dot{P}$, the evolution stalls and sources pile up at a characteristic period.

For typical magnetar parameters ($B_0 \sim 10^{14}-10^{15}$~G, $\tau_B \sim 10^4$~yr), this model predicts pile-up at periods of several seconds, qualitatively consistent with observations. However, as noted by PM02, sources near the maximum period (such as 1E~1841$-$045 with $P = 11.79$~s and $\dot{P} \sim 4.1 \times 10^{-11}$~s~s$^{-1}$) have large period derivatives, implying that field decay must be faster than dipole spin-down for these sources.

\citet{pon+13} considered a variant of the field decay model in which a highly resistive layer in the neutron star crust at densities corresponding to the nuclear pasta phases ($\rho \sim 10^{14}$~g~cm$^{-3}$) causes enhanced Ohmic dissipation of the magnetic field.
Their key insight is that the effective electrical resistivity in the pasta layer can be orders of magnitude higher than in the rest of the crust, due to electron scattering off the complex nuclear structures. This leads to a field decay timescale $\tau_{\rm Ohm} \sim 4\pi\sigma L^2 / c^2$,
where $\sigma$ is the electrical conductivity and $L$ is the characteristic length scale. \citet{pon+13} showed that this mechanism naturally limits spin periods to $\lesssim 12$~s: once the field decays below a critical value, the source becomes undetectable (either because X-ray luminosity drops or because the field becomes too weak to produce the characteristic magnetar emission). This model predicts that no magnetars should exist with $P > 12$~s, in agreement with our statistical constraints.

\citet{cha+00} and \citet{alp01} proposed a complementary scenario in which magnetars are surrounded by residual matter from their natal supernovae, a ``fallback disc''. In this picture, the period clustering results from the disc-magnetosphere interaction, which drives these systems towards torque equilibrium. Since the accretion torque is stronger than the magnetic dipole torques, the dipole magnetic fields required to explain spin-down rates are smaller. The defining magnetar bursts result from much stronger fields in the multipoles \citep{ert03,eks03}. Assuming that at a critical luminosity, corresponding to a critical accretion rate $\dot{M}_{\rm c} \sim 10^{13}$~g~s$^{-1}$ which in turn  corresponds to a critical temperature in the outer disc, the gas becomes too neutral for sustaining magnetorotational turbulence \citep{bal91,haw91,bal98}, turning into a ``dead disc.'' This equilibrium period at the critical mass accretion rate would be the maximum period for an object with a given magnetic field \citep{men+01stab}, and the cutoff period would represent the maximum period for objects with the largest dipole fields. The exact value of the critical luminosity depends on the strength of X-ray irradiation. The rapid spin-down rate of the magnetar with the largest period (i.e.\ 1E~1841$-$045 with $P=11.79$~s, $\dot{P}=4.1\times 10^{-11}$~s~s$^{-1}$) suggests that the transitions to the passive disc stages may occur before the systems reach torque equilibrium.

The period clustering of magnetars and XDINS has broader implications for our understanding of young neutron stars. The existence of a physical mechanism that terminates the X-ray active phase at $P \sim 16$~s implies:

\begin{enumerate}
    \item \textbf{A hidden population:} If magnetars continue to spin down beyond 12~s but become X-ray faint, there should be a population of very slowly rotating, highly magnetized neutron stars that are currently undetectable. 
    
    \item \textbf{Evolutionary connections:} The similar period distributions of magnetars and XDINS suggest possible evolutionary pathways connecting these populations. For example, magnetars with decaying fields might evolve into XDINS-like objects as their bursting activity ceases \citep{gul+14}. Within the fallback disc model, they could correspond to different initial conditions of the period and disc parameters \citep{alp01,gen24}.
    
    \item \textbf{Birth rates:} The observed period distribution, combined with assumptions about spin-down evolution, constrains the magnetar birth rate. If sources spend most of their active lifetime near the maximum period (as predicted for $n > 2$), the observed sample is representative of the steady-state population.
\end{enumerate}

It is instructive to ask why radio pulsars, despite having braking indices close to $n=3$, do not show a comparably tight upper cutoff in their period distribution. 
The answer lies in the nature of the termination mechanism. Radio pulsars cease emission when they cross the empirical \textit{death line} in the $P-\dot{P}$ diagram, corresponding to the condition that the electric potential drop across the polar cap becomes insufficient for pair creation. 
This death line is not at a fixed period but depends on $\dot{P}$ as well, so it does not impose a sharp upper cutoff in $P$ alone. 
Furthermore, since $\dot{P}\propto B^2/P$, radio pulsars span many decades in $\dot{P}$ at any given period, and their emission ceases over a wide range of periods. 
By contrast, magnetars and XDINS are powered by magnetic-field decay or by disc-magnetosphere torque equilibrium, mechanisms that impose characteristic timescales and, hence, characteristic periods at which the observable phase ends. 
This is the physical content of $\Pf$ in the PM02 framework: it is not a radio death line but a field-decay or torque-equilibrium cutoff that is approximately the same for all sources with similar initial fields.

We would like to point out a few caveats of the approach we followed in this work:

\begin{enumerate}

   \item \textbf{Constant braking index:} Our analysis assumes a constant braking index throughout the source lifetime. In reality, field decay, glitches, and magnetospheric changes can cause $n$ to vary substantially. The constant braking index model provides a phenomenological framework that captures the essential statistical properties of the period distribution without specifying the underlying physics. In a follow-up study, we will employ more physical models both for the isolated magnetar assumption and for the magnetars with fallback discs.
 
    \item \textbf{Single birth period:} We assume all sources are born with the same $\Pin$. A distribution of birth periods might broaden the constraints on $\Pin$, but not change $\Pf$ significantly.
    
   \item \textbf{Birth distributions.} Our analysis assumes specific initial conditions. A full population synthesis would require convolving our evolutionary tracks with realistic distributions of $P_0$ and $B_0$ \citep[see, e.g.,][]{gul+14}.
    
    \item \textbf{Selection effects:} Although PM02 argued that observational selection effects cannot account for the period clustering, systematic biases in detecting faint, long-period sources cannot be entirely excluded.
    
    \item \textbf{Steady-state assumption:} We assumed the magnetar population is in steady state. If the local population is dominated by sources from a recent burst of star formation (e.g., in the Gould Belt \citep{pop06}), the period distribution could be affected.
\end{enumerate}

Despite these limitations, the remarkable persistence of the 12~s cutoff of magnetars over two decades of discoveries, and the present cutoff proposed here for XDINS at about 15~s provides strong evidence that the period clustering reflects genuine astrophysics rather than statistical fluctuation or selection bias.


\section{Conclusions}
\label{sec:conclusions}

We have revisited the statistical analysis of magnetar period clustering originally performed by PM02, using the current sample of 30 magnetars and 8 XDINS. Our main conclusions are:

\begin{enumerate}
    \item The period clustering of magnetars and XDINS is genuine and statistically significant. For dipole spin-down ($n = 3$), the final (cutoff) period is constrained to $\Pf = 12.88 - 13.61$~s at 95 per cent confidence for the combined sample.
    
    \item The 12~s upper cutoff of the magnetar population has remained unchanged despite a threefold increase in the known magnetar population and the extension of the minimum period from $\sim 5$~s to $\sim 0.33$~s.
    
    \item The birth period is poorly constrained for $n \geq 2$, but becomes increasingly constrained to long values ($\Pin \sim$ few seconds) for $n < 2$.
    
    \item The XDINS period distribution is statistically consistent with the magnetar distribution, supporting evolutionary connections between these populations.
    
    \item The results can be qualitatively explained with physical models invoking magnetic field decay \citep{col+00}, crustal resistivity limits \citep{pon+13}, or fallback disc torque equilibrium \citep{alp01}. The scope of this work is limited to showing the existence of clustering in a rather model-independent way. To evaluate how well these models address the period-clustering phenomenon, one must perform a model-dependent analysis in which one employs the spin-down torques for each model separately and analyzes the posterior distributions. This is left for future publication elsewhere. 
\end{enumerate}

Future observations with current and next-generation X-ray observatories will continue to test these conclusions. The discovery of even a single magnetar with $P > 12$~s or an XDINS with 16 s would significantly impact the constraints on $\Pf$ values and challenge current theoretical models.

\section*{Acknowledgements}
I thank Mehmet Ali Alpar and Sinem Şaşmaz for their useful discussion and comments on the manuscript.

\bibliographystyle{mnras}
\bibliography{fallback,periods}

\bsp	
\label{lastpage}
\end{document}